\documentclass[conference]{IEEEtran}
\IEEEoverridecommandlockouts
\usepackage{cite}
\usepackage{amsmath,amssymb,amsfonts}
\usepackage{algorithmic}
\usepackage{graphicx}
\usepackage{textcomp}
\usepackage{xcolor}
\usepackage{url}
\usepackage{soul}
\usepackage{upgreek}

\def\BibTeX{{\rm B\kern-.05em{\sc i\kern-.025em b}\kern-.08em
    T\kern-.1667em\lower.7ex\hbox{E}\kern-.125emX}}
    
\begin{document}
\title{Blockchain-Enabled End-to-End Encryption for Instant Messaging Applications}

\author{\IEEEauthorblockN{Raman Singh}
\IEEEauthorblockA{\textit{School of Comp Sci \& Stats} \\
\textit{Trinity College Dublin} \\
Dublin, Ireland \\
\textit{Thapar Institute of Engineering \& Technology}\\
Patiala, India \\
raman.singh@thapar.edu}
\and
\IEEEauthorblockN{Ark Nandan Singh Chauhan}
\IEEEauthorblockA{\textit{School of Comp Sci \& Stats} \\
\textit{Trinity College Dublin}\\
Dublin, Ireland \\
chauhaar@tcd.ie}
\and
\IEEEauthorblockN{Hitesh Tewari}
\IEEEauthorblockA{\textit{School of Comp Sci \& Stats} \\
\textit{Trinity College Dublin}\\
Dublin, Ireland \\
htewari@tcd.ie}
}

\maketitle

\begin{abstract}
In this era of ubiquitous social media and messaging applications, users are becoming increasingly aware of the data privacy issues associated with such apps. Major messaging applications are moving towards end-to-end encryption (E2EE) to give their users the privacy they are demanding. However the current security mechanisms employed by different service providers are not unfeigned E2EE implementations, and are blended with many vulnerabilities. At present, the major part of the E2EE mechanism is controlled by the service provider's servers, and the decryption keys are also stored by them in case of backup restoration. These shortcomings diminish user confidence in the privacy of their data when using these apps. A public key infrastructure (PKI) can be used to circumvent some of these issues, but it comes with high monetary costs, which makes it impossible to roll out on a global scale. This paper proposes a blockchain-based E2EE framework that can mitigate many of the contemporary vulnerabilities in today's messaging applications. A user's device generates the public/private key pair during application installation, and asks its mobile network operator (MNO) to issue a digital certificate and store it on a public blockchain. Any user can fetch a certificate for another user from the application server, and communicate securely with them using a ratchet forward encryption mechanism.
\end{abstract}

\begin{figure*}[!htb]
\centering
\includegraphics[width=7.0in]{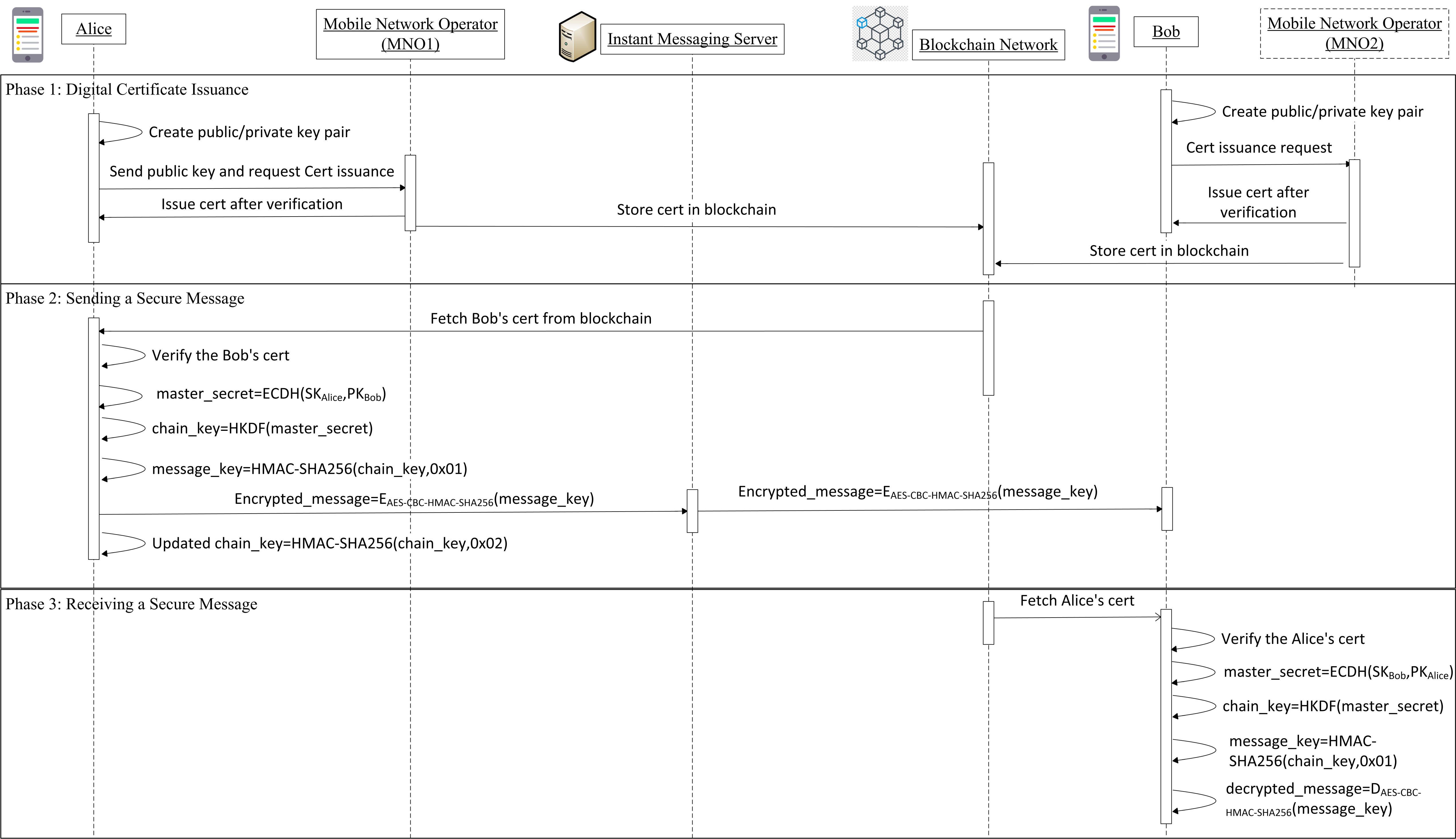}
\caption{Sequence Diagram Illustrating Various Phases}
\label{fig:figure1}
\end{figure*}

\section{Introduction}
``Data is the new oil" and big technology companies are using every tool at their disposal to store and utilize user data for commercial gain, as the personal and behavioral data of their users is worth millions of dollars to them if mined to its maximum potential. For example, a simple message exchange with a loved one or a friend about a visit to a cafe or a review of a local fashion store can generate significant insights for big technology companies, and can be sold to third parties, given the large number of users on their platforms. However, users are slowly becoming aware of privacy issues associated with these apps, and are increasingly wary of being listened to or of their actions being monitored by these technology companies. The majority of service providers do not charge an upfront fee for the services provided by them, but openly make use of user data for generating more profit than the underlying service costs. The saying ``if you are not paying for the product, then you are the product" signifies these free services, where companies make revenue by advertising products to you. Many users willingly share personal information or documents like photos, driver licenses, passports, bank account details, etc. through these messaging apps, that make them vulnerable to identity theft or other cybersecurity attacks. There is also a chance of leakage of sensitive information like business dealings, intellectual property, non-disclosure agreements, etc. For all of the above reasons, the fundamental right of privacy is prime and should be provided for all online communications.

The latest update of user policies by WhatsApp \cite{whatsapp1} once again highlights the awareness of privacy issues in online communications. Despite WhatsApp's claim of end-to-end encryption (E2EE), users fear the misuse of the data collected by the company and its associated organizations. To protect their privacy, many users have started migrating to the more secure Signal App \cite{signalapp1}. Both these apps make use of Signal's security mechanism, but the open-source nature of the Signal App provides more confidence to its users, whereas WhatsApp's implementation is proprietary and not open to scrutiny by independent third parties. Similarly, the roll out of 5G networks around the world is slowing because of the backlash faced by Huawei \cite{huawei1}, due to doubts circulated about privacy violations by its 5G equipment. It has been alleged by intelligence agencies around the world that Huawei's 5G equipment can capture plaintext data while en-route to a destination, which violates the requirement of E2EE for data communications. Whether these claims are true or not still remains to be proven. However the one thing these rumours have done is to dent user confidence in these technologies.

We believe that a global public key infrastructure (PKI) is the answer to these privacy issues. However the current PKI model does not lend itself well to large-scale deployment, primarily because of the cost involved in issuing and maintaining digital certificates for the many millions of users. To resolve the PKI deployment issue, in this paper we propose a blockchain-enabled E2EE framework that can provide real end-to-end encryption for online communications. The blockchain makes it possible to implement the large-scale PKI system virtually at ``zero cost" \cite{X509Cloud}. Unlike WhatsApp, the proposed framework never allows the server to store any keys, or to participate in the encryption/decryption process. The public/private keys are generated by users and they never share the private key with anyone. The validity of digital certificates is maintained by the blockchain mechanism, and no one else other than the sender/receiver can read the message.

\section{Fragility of Secure Messaging Apps}
Many popular messaging applications like WhatsApp now provide E2EE channels for their users. The WhatsApp and Signal apps use the same Signal protocol \cite{signalapp2} for key exchange and encryption, but use different messaging protocols. At install time the application client creates a number of keys and sends them to the WhatsApp server, where the server stores the keys against a user identifier \cite{whatsapp2}. WhatsApp has the option of verifying the public keys of users, but the mechanism employed by it are not robust and have serious vulnerabilities from session hijacking. Apart from that, there is no trusted third-party involvement to verify the validity of keys stored on the WhatsApp servers.

Users have no option but to trust that the public keys provided by the WhatsApp server are in no way altered on the WhatsApp databases or by a man-in-the-middle (MITM) attack. The design of WhatsApp also makes it possible for attackers to infiltrate a messaging group. Attackers then have the liberty to modify the information of the group as it is not encrypted, and hence chances of becoming a group member exist. An attacker can also silently drop messages and send an acknowledgment to the sender that gives the impression that the receiver received the message \cite{rosler2018more}. In recent times WhatsApp has been keen to update its security policies. The new security policies allow it to share user details like account information, user connections, transaction/payment-related data, usage/log information, location information, cookies, etc. with parent/sister companies such as Facebook \cite{whatsapp_security}. The location of a user can be accessed by governments and other organizations easily because WhatsApp now stores this information on its servers.

For a business customer who is using a WhatsApp Business Account and Facebook's secure hosting infrastructure, the messaging protocol is different from the free user accounts. When Facebook acts as a hosting provider to a business, it will use the messages it processes on behalf of the business and at the instruction of the business. Thus the messages cannot be considered end-to-end encrypted as they can be accessed by Facebook and other businesses for marketing purposes/advertising on Facebook. This means that if a non-business WhatsApp user interacts with a business account their chats cannot be considered end-to-end encrypted.

The way WhatsApp handles the backup mechanism does not provide true end-to-end encryption. The backup copy is stored on the cloud of the user based on their operating system such as Google Drive, OneDrive, iCloud, etc., but the decryption key is stored on the WhatsApp servers. Whenever a user wants to restore a backup, the WhatsApp server sends the user's decryption key to his/her device. This mechanism poses a serious vulnerability to the confidentiality of the messages stored in the backup. On the one hand, the WhatsApp server will always have the liberty to decrypt messages, while on the other hand hackers can copy the backup and trick the WhatsApp server to send the decryption key to them.

Other popular messaging applications like Signal and Threema \cite{threema1} have a similar issues when it comes to  end-to-end encryption. Since every group member in Signal has administrative privileges, hackers will have the liberty to contribute to the group messaging which makes infiltrating the group possible. The Signal protocol is designed to provide a mechanism to detect which messages are not received by the recipient, but this is not effective, and messages can still be secretly dropped during the communication. It is also possible for malicious users to re-order the received messages by manipulating the server. Researchers have demonstrated breaches in security in the Threema application. They found that perfect forward secrecy, future secrecy, or traceable delivery is not achievable in the application. The researchers also proved that the messaging application can be used to resend the messages without detection of duplicate messages. The application orders the received messages using receiving time and the sending time is not protected on the end-to-end layer. Therefore, it is possible for the malicious users to arbitrarily reorder the messages during communication \cite{rosler2018more}.

\section{Blockchain-Enabled E2EE for Instant Messaging}
Our proposed blockchain-backed E2EE mechanism aims to provide real confidentiality without any fear of the sharing secret keys with third parties like instant messaging (IM) servers. The secure messaging mechanism spans across many entities such as mobile network operator (MNO), blockchain nodes, IM servers, sender and receiver devices. Fig. \ref{fig:figure1} details the sequence diagram of the proposed security framework for these phases. The figure also details the various cryptographic key generation procedures required by our system.  We make use of a ``permissioned" blockchain with entities such as the MNOs and IM servers authorized to write to the blockchain via their attached blockchain node.
\newline

\begin{figure}
\centering
\includegraphics[width=3.5in]{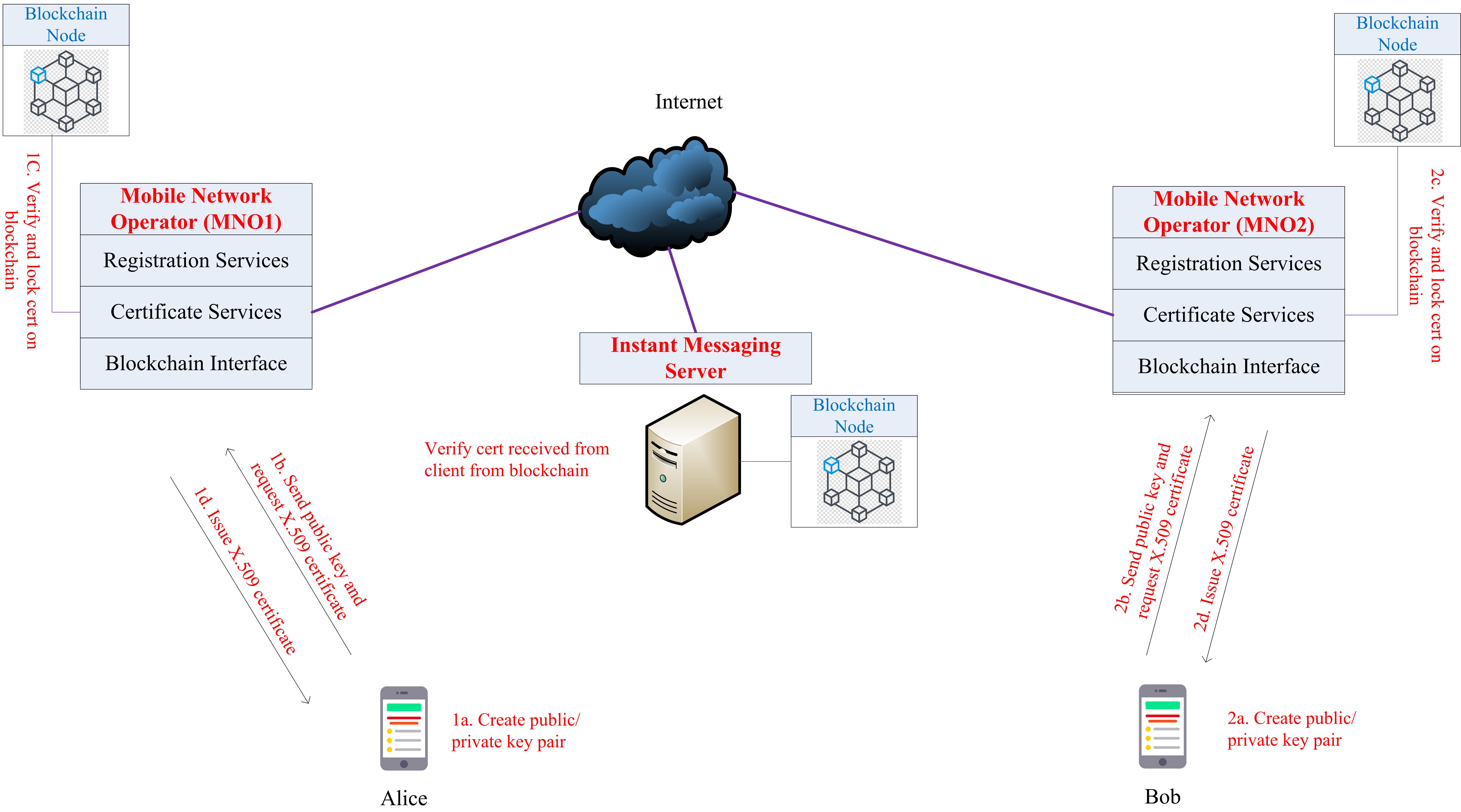}
\caption{Certificate Generation and Registration}
\label{fig:figure2}
\end{figure}

\textbf{Phase 1 - Certificate Generation and Registration:} 
The process starts with the registration of users with the IM server as shown in Fig. \ref{fig:figure2}. A user downloads the messaging app and while installing it creates an asymmetric public and private key pair. The public key is shared with associated MNO that in turns verifies the public key and issue a digital certificate to the user. The same certificate is also stored on the blockchain node connected to the MNO. The MNOs in our system are a trusted third-party (TTP). The IM server is also connected to a blockchain node and has access to stored digital certificates using a users' unique identifier.

\begin{figure*}[!htb]
\centering
\includegraphics[width=7.0in]{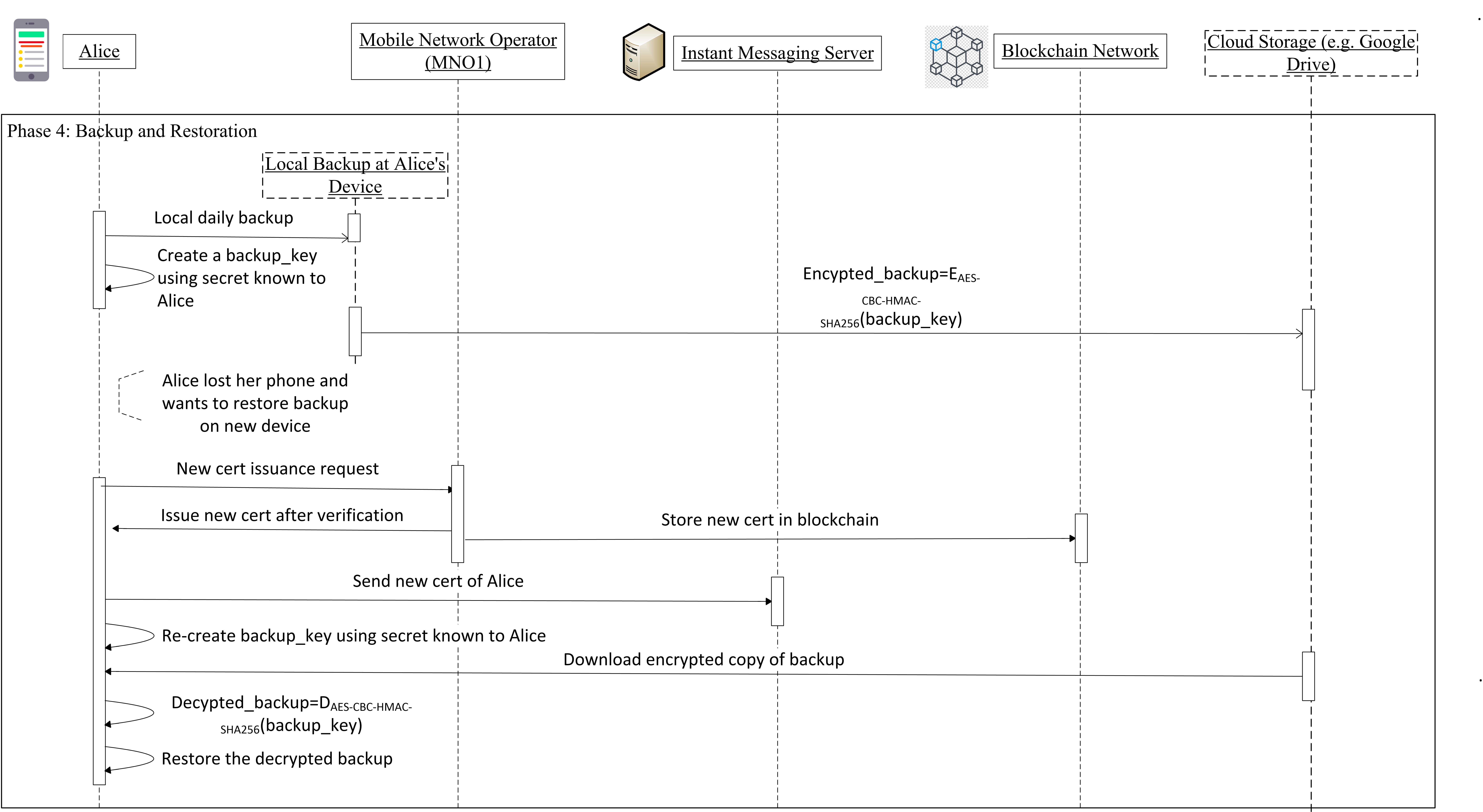}
\caption{Sequence Diagram Illustrating Backup and Restoration}
\label{fig:figure3}
\end{figure*} 

In the proposed framework, Alice needs to fetch Bob's certificate from the IM server before she sends messages to him and vice versa. The IM server and MNO both have the option to verify the authenticity of the certificates stored on the blockchain node. MNOs also provide a mechanism for users to access the blockchain node and verify the certificates of receivers. The one problem with the present X.509 based PKI system is with efficient and time-bound revocation of issued certificates. In order to revoke an issued certificate, the system needs to enter it in Certificate Revocation List (CRL) and the CRL should be propagated widely within the system. The window between actual entry in CRL and its subsequent update with each entity may invoke misuse of the revoked certificate. This problem can be solved using blockchain-backed PKI as the virtue of its functionality, only the latest transaction will be fetched first. It means, if an authenticated entity wants to revoke any certificate, it will just add a dummy certificate on the blockchain and this ``dummy" certificate will be fetched first, telling the fetching entity that the particular certificate is no longer valid. In this way, there will be no need to prepare lengthy CRL \cite{X509Cloud}.   
\newline

\textbf{Phase 2 - Sending a Secure Message:}
Alice's messaging app generates the shared master secret key using the secret key of Alice and the public key of Bob. A \textit{chain\_key} is then created using a hash based key derivation function (HKDF), and a \textit{message\_key} is then derived from these keys. The \textit{chain\_key} is regularly updated using the ratchet forward method and the sender encrypts messages using the dynamic \textit{message\_key}. If Bob does not reply to Alice's messages, in this case, Alice will generate a new \textit{chain\_key}, encrypt the message using the newly created \textit{message\_key} and again update the \textit{chain\_key}.

The ratchet forward mechanism \cite{ratchet} updates the \textit{chain\_key} for every new message thus providing forward secrecy to the system architecture. This ensures that if any single session key gets compromised, the rest of the data on the system remains safe. Only the data protected by the compromised key is vulnerable. Forward secrecy ensures that if the current key is exposed to an attacker, the attacker cannot read the messages of any previous sessions.

The Signal protocol also provides end-to-end encryption using the ratchet protocol along with various cryptographic elements like HKDF, symmetric key cryptography and Elliptic-curve Diffie–Hellman (ECDH) key exchanges. ECDH is a key agreement protocol which helps two different parties to establish a shared secret over an insecure channel. These two parties should have an elliptic curve public-private key pair, and the shared secret is used to compute a symmetric key. To generate the same symmetric key, sender and receiver both should agree on elliptical curve type and key size beforehand. Compared to other public-key algorithms, ECDH offers faster computational speeds and suitable smaller key sizes for the same level of security \cite{karbasi2021singleton}.
\newline

\textbf{Phase 3 - Receiving a Secure Message:}
At the receiver side, the messaging app can generate the \textit{message\_key} using the same mechanism ad continue to decrypt the received ciphertext. The \textit{message\_key} is generated from the \textit{chain\_key} and the \textit{chain\_key} will be updated after that. No shared secret key is stored on the IM server or shared with the receiver eliminating any vulnerability arising due to it. The similar framework can be extended to a proper decentralized version of a secure messaging application.

In this version, no intermediate IM server is required. The sender can fetch the certificate of a receiver from the blockchain associated with the MNO. The encrypted messages are then forwarded to the receiver directly without involving any IM server in-between. This kind of setup is more suitable for organization-to-organization messaging rather than individual users, because all users may not be online at all times and hence may drop the intended messages.
\newline

\textbf{Phase 4 - Backup and Restoration:}
This phase talks about backup and restoration of user data. Unlike WhatsApp where the backup decryption key is stored on the application server, our proposed framework advises users to generate their backup decryption key using a known secret (e.g. a password). The backup data is stored on the user's Google Drive/iCloud/OneDrive etc. and encrypted by the \textit{backup\_key}. When the user changes their device, they can download their backups from the cloud drive and decrypt them using their \textit{backup\_key} generated from a known secret. Fig. \ref{fig:figure3} depicts the sequence diagram for the backup and restoration phases. Since the \textit{backup\_key} is generated using the known secret, the user can re-generate it using the same process with the same known secret. If the user loses his/her \textit{backup\_key} and also forgets the known secret there is no way to recreate the key.

\section{Results Analysis}
The IM server in our proposed framework is implemented using Google Firebase \cite{firebase1}, whereas the messaging app with features like login, registration, one-to-one IM, group IM, backup, and certificate verification is developed using an Android application. To implement the blockchain functionality, the Ethereum \cite{ethereumofficial} blockchain is implemented on the Docker \cite{dockerofficial} platform. Our analysis tries to determine the performance of the AES256 in CBC mode for encryption and decryption, HMAC with SHA256 for message verification code, and the total time taken for the E2EE process on the Android emulators. Based on the string input of varying lengths, the time for various processes within the application has been measured. The experiments were conducted using the Google Pixel 4 XL emulator using Google APIs Intel Atom(x86) system image with 1536Mb of RAM and 384Mb heap size.

\begin{figure}
\centering
\includegraphics[width=3.3in]{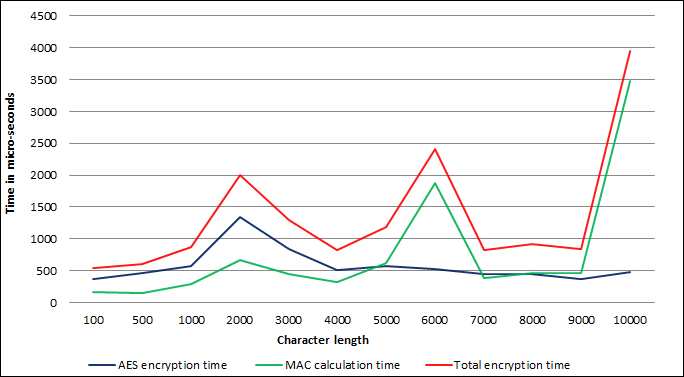}
\caption{Encryption Time Analysis}
\label{fig:figure4}
\end{figure}

Fig. \ref{fig:figure4} shows the graph for three processes: AES encryption time, MAC calculation time, and the Total encryption time in seconds against string inputs of up to 10,000 characters. Using the data gathered from the graph, the average AES encryption time was found out to be 576.508$\upmu$s, the average MAC calculation time was 778.1958$\upmu$s, and the average Total encryption time was 1354.859$\upmu$s. The graph reveals a generally linear relationship between the amount of time taken for AES encryption in CBC mode with increasing character lengths. The results show that a total encryption time of 3995.663$\upmu$s is taken when the character length is increased to 10,000. The results also show that AES encryption time is almost stable with increasing character length, but MAC calculation time varies with different character lengths.

\begin{figure}
\centering
\includegraphics[width=3.3in]{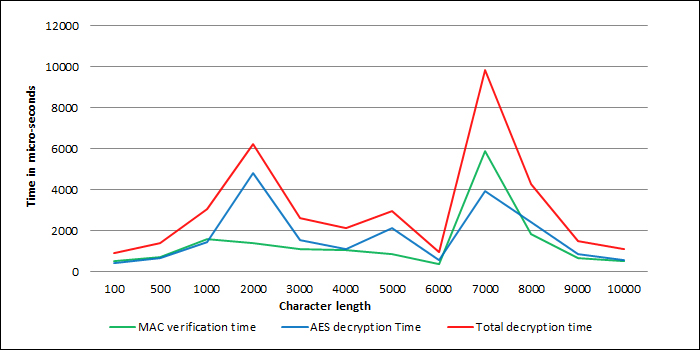}
\caption{Decryption Time Analysis}
\label{fig:figure5}
\end{figure}

Fig. \ref{fig:figure5}, shows the graph for decryption time analysis for calculating the AES-CBC decryption time, MAC verification time, and the Total decryption time. The average AES decryption time was found out to be 1699.179$\upmu$s, the average MAC verification time was 1370.069$\upmu$s, total average decryption time was 3069.248 micro-seconds. The results show that total decryption time is stable to a large extent with increasing character lengths.

\section{Conclusions and Future Work}
With the lack of worldwide implementation of a public key infrastructure, existing messaging applications are providing end-to-end encryption using their own servers as a trusted entities. Backup decryption keys are also stored on the IM server and the end-to-end encryption/decryption mechanism is controlled by service providers themselves. The encryption keys are stored on the organization’s server if users are using business messaging services. The current messaging services are also vulnerable to being accessed in a group by unauthorized users. The primary reason for these vulnerabilities is the absence of trusted third-party authentication of users and IM servers.

Blockchain technology can be used to overcome these issues, and it can provide a real end-to-end encrypted messaging service. A mobile user creates public/private key pair on their device during the application installation phase, and a digital certificate is created by the mobile network operator (a trusted third-party) for that user, based on information provided by them. The public-key digital certificate is then stored on the blockchain and the mobile user sends this to the IM server. The same process is followed by each user and the IM server can verify any user’s certificate from the blockchain network. A sender can fetch the receiver‘s digital certificate from the blockchain before she encrypts a message. 

Our proposed framework does not rely on an IM server for encryption/decryption and hence can be treated as a real end-to-end encryption mechanism. Currently we are working on two other important aspects of the proposed framework, the first is for providing secure ``group messaging" while the second is for testing the ``scalability" of the framework in a blockchain-based environment. The group messaging protocol makes use of one-to-one encrypted channels for key exchange between users. Once a group is created, the administrator of the group generates a \textit{group\_key} on their device, and shares the \textit{group\_key} with each new member over a one-to-one encrypted channel. Messages sent by a group member are broadcast by the IM server in a fan-out fashion. Group messages can be encrypted using the \textit{message\_key} derived from the \textit{group\_key} with AES256 in CBC mode and HMAC-SHA256 for authentication. We will also test the proposed framework against the scalability of a blockchain with a large number of transactions and blocks, in terms of certificate fetching time, certificate verification time, new certificate issuance time and the overall efficiency of the system.


\begin{thebibliography}{10}

\bibitem{whatsapp1} 
Simple. Secure, Reliable messaging, https://www.whatsapp.com/?lang=en,  Accessed on 08-03-2021.

\bibitem{signalapp1} 
Signal Private Messenger, https://signal.org/en/, Accessed on 08-03-2021.

\bibitem{huawei1} 
Huawei, https://www.huawei.com/en/, Accessed on 08-03-2021.

\bibitem{X509Cloud}
H. Tewari, A. Hughes, S. Weber and T. Barry, ``X509Cloud — Framework for a ubiquitous PKI" MILCOM 2017 - 2017 IEEE Military Communications Conference (MILCOM), Baltimore, MD, USA, 2017, pp. 225-230.

\bibitem{signalapp2}
Signal Technical Specification, https://signal.org/docs/, Accessed on 12-04-2021.

\bibitem{whatsapp2} 
Technical White Paper: WhatsApp Encryption Overview, http://www.cdn.whatsapp.net/security/WhatsApp-Security-Whitepaper.pd, Accessed on 23-02-2021. 

\bibitem{rosler2018more}
P. Rösler, C. Mainka and J. Schwenk, ``More is less: On the end-to-end security of group chats in Signal, WhatsApp, and Threema", IEEE European Symposium on Security and Privacy (EuroS\&P), pp. 415-429.

\bibitem{whatsapp_security}
https://faq.whatsapp.com/general/security-and-privacy/were-updating-our-terms-and-privacy-policy/, Accessed on 09-03-2021.

\bibitem{threema1}
The messenger that puts security and privacy first, https://threema.ch/en, Accessed on 08-03-2021.

\bibitem{ratchet}
https://nfil.dev/coding/encryption/python/double-ratchet-example/, Accessed on 11-03-2021.

\bibitem{karbasi2021singleton}
A.H. Karbasi  and S. Shahpasand, ''SINGLETON: A lightweight and secure end-to-end encryption protocol for the sensor networks in the Internet of Things based on cryptographic ratchets'' The Journal of Supercomputing, 77(4), pp.3516-3554.

\bibitem{firebase1}
Firebase helps you build and run successful apps, 
https://firebase.google.com/, Accessed on 13-07-2021.

\bibitem{ethereumofficial} 
Ethereum, https://ethereum.org/en, Accessed on 20-07-2021.

\bibitem{dockerofficial}
Docker Container, https://www.docker.com, Accessed on 20-07-2021.
\end{thebibliography}
\end{document}